\documentclass[twocolumn,showpacs,preprintnumbers,amsmath,amssymb,superscriptaddress]{revtex4}

\usepackage{graphicx}
\usepackage{dcolumn}
\usepackage{bm}

\begin{document}

\preprint{DFI070301-v.2.6}

\title{Demonstration of displacement- and frequency-noise free laser interferometry 
using bi-directional Mach-Zehnder interferometers}

\author{Shuichi Sato}
\email{sato.shuichi@nao.ac.jp}
\affiliation{
TAMA project, National Astronomical Observatory of Japan,
\\2-21-1, Mitaka, Osawa, Tokyo 181-8588 Japan}

\author{Keiko Kokeyama}
\affiliation{
The Graduate School of Humanities and Sciences, Ochanomizu University,
\\2-1-1, Otsuka, Bunkyo-ku, Tokyo 112-8610 Japan}

\author{Robert L. Ward} 
\affiliation{
LIGO Project 18-34, California Institute of Technology, Pasadena, California 91125, USA}

\author{Seiji Kawamura}
\affiliation{
TAMA project, National Astronomical Observatory of Japan,
\\2-21-1, Mitaka, Osawa, Tokyo 181-8588 Japan}

\author{Yanbei Chen}
\affiliation{
Max-Planck-Institut f\"ur Gravitationsphysik, Am M\"uhlenberg 1, 14476 Potsdam, Germany}
\author{Archana Pai}
\affiliation{
Max-Planck-Institut f\"ur Gravitationsphysik, Am M\"uhlenberg 1, 14476 Potsdam, Germany}
\author{Kentaro Somiya}
\affiliation{
Max-Planck-Institut f\"ur Gravitationsphysik, Am M\"uhlenberg 1, 14476 Potsdam, Germany}

\date{\today}

\begin{abstract}
We have demonstrated displacement- and frequency-noise free laser interferometry (DFI) by 
partially implementing a recently proposed optical configuration using bi-directional 
Mach-Zehnder interferometers (MZIs).  This partial implementation, the minimum necessary
to be called DFI, has confirmed the essential feature of DFI: the combination of two MZI signals 
can be carried out in a way that cancels displacement noise of the mirrors while maintaining gravitational wave signals.
The attained maximum displacement noise suppression was 45\,dB.
\end{abstract}

\maketitle


Gravitational waves are ripples of space-time curvature. For an array of test masses separated from each other by a length scale $\sim L$, that is not too much bigger than the gravitational wavelength, the effect of gravitational waves can often be thought of as creating relative test-mass motions with order $\sim L h$, or creating additional time delays of order $L h/c$ for light pulses traveling between these test masses. One might then think that, in order to measure gravitational waves, we need to make sure that (i) our clocks must have timing errors lower than $\delta t \sim L h/c$, and that (ii) the test masses must follow geodesics within an accuracy of $\delta x \sim Lh$.  In fact, {\it both are unnecessary.}  Point (i) was shown to be unnecessary more than a century ago, with the invention of  {\it interferometry} --- in which one light beam is split into two beams, each beam then measures the same quantity but with opposite sign, and then the two beams are subtracted coherently to eliminate the {\it common} fluctuations in light frequency while retaining the {\it differential} signal.  An extended application of this idea in gravitational-wave detection is the so-called Time-Delay Interferometry (TDI)~\cite{TDI}. Point (ii) is more specific to the problem of gravitational-wave detection, and was addressed only recently. As was shown theoretically in Refs.~\cite{KC1,KC2}, when each test mass in an $N$-test-mass array sends and receives light pulses from all other test masses, and if the array contains enough test masses [$N > (d+2)$, where $d$ is the number of spatial dimensions of the array], then there exist combinations of time delays (i.e., signals) which do not sense test-mass motions or timing noises, but do (usually) sense gravitational waves. The existence of such noise-free signals relies on the fact that the gravitational wave contribution to light pulse timing delays takes a form different from that of test-mass motions --- we therefore can combine the pulse time delays in such a way as to eliminate the {\it common} test-mass motions, while retaining {\it differential} gravitational-wave signals.  We will call configurations that cancel both timing and displacement noises Displacement-noise-Free Interferometry (DFI). 

Recently, practical optical designs of DFI using laser interferometry in 2- and 3-dimensions have been proposed~\cite{KC3}. In these configurations, the conventional, equal-arm Mach-Zehnder interferometer was used as a building block to eliminate laser noise.  Four such Mach-Zehnder interferometers were combined, in such a way that they form two pairs of counter-propagating  Mach-Zehnders. Within each pair, the two Mach-Zehnders share the same beamsplitters and folding mirrors; subtraction of their outputs balances out displacement noise from motions of the folding mirrors.  The two pairs share the same beamsplitters, which allows the elimination of beamsplitter displacement noise.  It should be mentioned that DFI configurations only work for non-zero frequencies, because at nearly zero frequency gravitational waves are indistinguishable from relative mirror motions.  For configurations studied so far (without optical cavities), the transfer function from $h$ to effective phase shift is $(\omega Lh/c) (\Omega L/c)^{\alpha}$ with $\omega$ the optical frequency, $\Omega$ the gravitational-wave frequency, and  $\alpha=2$ for 3-D configurations, and $\alpha=3$ for 2-D configurations.  This means we need to observe at frequencies near $c/L$ in order to have a transfer function comparable to conventional configurations. 
For instance, for the ground-based interferometers with arm lengths limited to several kilometers,
$c/L$ becomes around 100\,kHz, so it seems difficult for current DFI configurations to compete with  them, which use optical cavities to increase the effective arm length targeting at gravitational waves at below kHz frequencies.
Nevertheless, we speculate that it will eventually find realistic applications in future detectors, or at least 
become one factor to consider during the invention of new detector configurations.
Another thing which should also be mentioned is that DFI can cancel out frequency- and displacement-noise 
of the component optics completely, however, any practical noises other than these, such as an intensity 
noise of the laser, noise arising from the scattered light and electronic noise are potential noise sources  in the DFI signal.
Therefore, in practical application of DFI, these noises should be carefully suppressed to low enough levels.

In this paper, we study DFI experimentally using a single pair of counter-propagating Mach-Zehnders, demonstrating the elimination of folding mirror displacement noise and the response to gravitational waves.  We use lab-scale devices, with $L \sim $ 1 meter, which indicates a frequency scale of $\sim 100\,$MHz.  Due to this high operation frequency, we simulate motions of folding mirrors with the phase shift induced by an EOM (Electro-Optical Modulator) located at the position of the folding mirror.  We place another EOM systematically at various other locations on the optical path, and use the phase shift it induces to map out the {\it Green's Function} of the device's gravitational-wave response.  We can do so because general relativity predicts that the effect of gravitational waves on our interferometers is equivalent to that of a medium with time- and location-dependent refractive index, and because we do not intend to test general relativity in this experiment. In order to see this, we write the phase shift imposed by a gravitational wave (in the so-called Transverse-Traceless, or TT gauge) on a light beam emitted from $\mathbf{x}_0$ with spatial direction $\mathbf{N}$:
\begin{equation}
\phi^{\rm gw}(t) = \frac{\omega}{c} \int_0^L N_i N_j   h_{ij}^{\rm TT}(t + l/c,\mathbf{x}_0 + \mathbf{N} l) dl\,.
\end{equation}
Here $t$ is the emission time, $h_{ij}^{TT}(t, \mathbf{x})$ is the gravitational-wave metric perturbation, see Eq.~(6) of Ref.~\cite{KC3}. Except at places where the beams intersect (negligible), we have a unique $\mathbf{N}$ vector, up to a sign difference (because we have counter-propagating beams overlapped with each other), at all points the light might travel through. This leads to a truly unique $N_i N_j$ at each point, and the effect of the gravitational wave is the same as that of a medium with the following refractive-index distribution:  
\begin{equation}
n^{\rm gw}(t,\mathbf{x}) =  N_i(\mathbf{x}) N_j(\mathbf{x}) h_{ ij}^{\rm TT}(t,\mathbf{x})\,.
\end{equation}

	\begin{figure}[ht]
	\begin{center}
	\includegraphics[width=8.5cm]{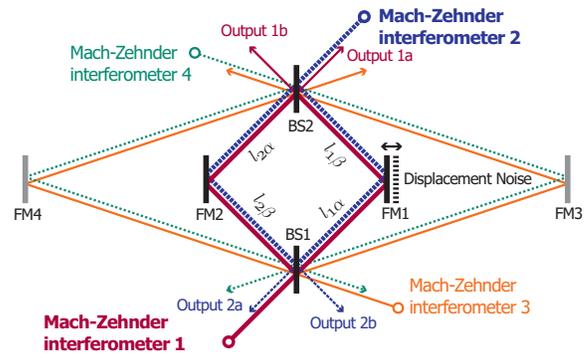}
	\caption{\label{fig1:schematic} (color online) 
	Schematic of one possible 2-D optical configuration of a displacement-noise-free interferometer.
	Light beams which follow common paths (thick, thin) 
	travel in opposite directions, and the four MZIs share the two beam-splitters (BS1 and BS2). The motion of each folding
	mirror (FM) is sensed by two MZIs, and the motion of the beam-splitters is sensed by all four MZIs.}
	\end{center}
	\end{figure}	
As depicted in Fig.~\ref{fig1:schematic}, the full 2-D DFI configuration (essentially a {\it squished} version
of the 3-D configuration in Ref.~\cite{KC3}) is composed of two pairs of counter-propagating MZIs, for a total
of four MZIs.  Mach-Zehnder interferometers of equal arm length operated on the dark fringe are insensitive to laser frequency
noise, and by super-imposing two counter-propagating Mach-Zehnders on the same optical path, the displacements of
the folding mirrors are sensed redundantly (once by each direction of the MZI) and can thus be unambiguously 
removed from the signal.  An additional pair of counter-propagating MZIs can then be added, which shares beamsplitters with the first
pair--this second pair allows redundant sensing of beamsplitter displacements.  As discussed in \cite{KC3},
the signals from the four MZIs can be combined in such a way as to cancel the displacements of all the optics
while retaining sensitivity to gravitational radiation.

For this experiment we have constructed a partial-DFI composed of a single pair of counter-propagating MZIs to 
demonstrate the cancellation properties of the DFI.  In Fig.~\ref{fig1:schematic}, the inner pair of MZIs, composed 
of BS1, BS2, FM1, FM2, has equal-length arms with folding mirrors located at the mid-point of each arm.  As the 
folding mirrors experience some displacement noise, the light field incident on those mirrors is phase modulated.
Because the folding mirrors are at the mid-point of the arms, and there is light traveling in both directions in
the MZI, these displacement noise based phase modulations will arrive at output 1 and output 2 simultaneously.  
Direct subtraction of the outputs 1 and 2 then allows one to cancel the motion of the folding mirrors.  Any signals
which phase modulate the light and which do not arise at the midpoint will not arrive simultaneously at the outputs,
and thus will not be fully cancelled. 

The transfer function of the displacement motion of folding mirror (FM1) to the signal port 
is 
\begin{eqnarray}
\label{d-res}
H_{D1(a,b)}=\pm(\omega/c)\exp(-{\rm i} l_{1\beta}\Omega/c)\\
H_{D2(a,b)}=\pm(\omega/c)\exp(-{\rm i} l_{1\alpha}\Omega/c)
\end{eqnarray}
where $c$ is the speed of light, $\omega$ is the laser frequency and $\Omega$ is a fourier frequency 
of the displacement motion.
The signals from output ports $n$a and $n$b have opposite sign for phase variation, 
depending on the fringe condition where the MZIs are controlled.
When the folding mirror is located at the exact center of the MZI arm (suppose 
$l_{1\alpha}=l_{1\beta}=l$), all four signals display an identical
frequency dependence in their transfer function (with different sign) as
\begin{equation}
H_{D(1,2)(a,b)}=\pm(\omega/c)\exp(-{\rm i} l\Omega/c)
\end{equation}
which enables the cancellation of signals due to displacement noise.

In contrast, the frequency response to the gw signal in a simplest case is given as 
\begin{eqnarray}
H_{gw(1,2)}=\pm(4\omega/{\rm i}\Omega)\exp(-{\rm i} l\Omega/c)\sin^{2}{(l\Omega/2c)}
\end{eqnarray}
for symmetric MZIs ($l_{n\alpha}=l_{n\beta}=l$), supposing normal incidence of gws 
with most effective polarization.
As a substitute for exact simulation of the gw signals, one can generate a phase modulation
signal at a location that is not the mid-point of an arm, and thus map the response of the 
interferometer to non-symmetrically located signals, by observing the difference in arrival times
between the two detection ports.

The response function to such a signal is as given in 
Eq.\eqref{d-res}.
So, as a result, the signal combination of MZI\,1 and MZI\,2 is given as 
\begin{eqnarray}
\label{sgw-res}
H_{sgw}=\pm(\omega/c)(\exp(-{\rm i} l_{1\alpha}\Omega/c)-\exp(-{\rm i} l_{1\beta}\Omega/c))
\end{eqnarray}
which shows complete cancellation of the displacement motion of the folding mirror, while allowing
{\it simulated}-gw signals to remain.

\begin{figure}[ht]
	\begin{center}
	\includegraphics[width=8cm]{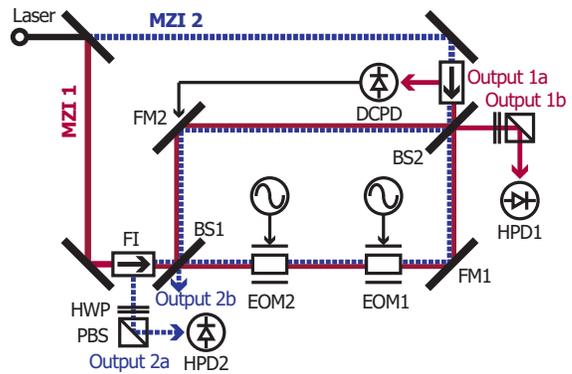}
	\caption{\label{fig2:setup}  (color online) 
	The experimental setup of partial DFI. MZI\,1 and MZI\,2 share a common laser source and 
	Mach-Zehnder interferometer parts. The scale of the MZI was about 4\,m, for which purpose 
	the optical paths of the MZI arms had to be folded several times on the optical table, which 
	does not affect the DFI features in this experiment.}
	\end{center}
	\end{figure}

The practical experimental setup is shown in Fig.~\ref{fig2:setup}.
The laser source is a commercial solid-state Nd:YAG laser (Lightwave Model-126) yielding 
100\,mW at 1064\,nm.
The output beam is split into two, and each beam introduced to two MZIs after passing 
through Faraday isolators (FI), which allow detection of light exiting the counter-propagating
MZI.
The light paths of both MZIs are carefully adjusted and superposed on each other so that 
the two MZIs can exactly share the displacement noise and the effect of {\it simulated}-gw signals.
The arm length of the MZIs affects the frequency where the MZI response to gw signal 
is maximized; in this experiment, the arm lengths were chosen to be around 3.6\,m, to lower the 
peak frequency to several tens of MHz.
To enable high frequency operation, the phase modulation due to both displacement noise and a gw signal were simulated using electro-optic phase modulators
located on one of the MZI arms.
The displacement noise simulator (EOM1) should be located at the exact center of the arm, 
so the position of EOM1 was carefully tuned, while the gw simulator (EOM2) was placed roughly
11cm away from the BS1.
The location of the EOM2 also determines the frequency of maximum sensitivity of DFI to 
the {\it simulated}-gw signal: a more asymmetrically placed EOM will give a lower peak frequency.

One of the output ports of MZI\,1 (output 1a) was monitored with a DC-detector (DCPD) and an error signal created
by subtracting a static offset; this signal was fed back to a PZT-actuated folding mirror (FM2) after appropriate filtering to 
give a mid-fringe locking control.
The control bandwidth was very low, around 400~Hz, so that the simulated signal by EOMs 
would not be suppressed by the fringe control feed back loop in the higher, more interesting frequency band.
Once the fringe of the MZI\,1 is controlled, that of MZI\,2 is also automatically controlled 
because the two MZIs share common optical paths.

Other output lights (output 1b and 2a) were received with high-speed photo detectors 
(New Focus 1611-AC) to monitor differential optical path length variations.
The polarization optics (HWP:\,half wave plate and PBS:\,polarizing beam splitter) just in front of both high speed detectors 
act as an optical attenuator, which serves as a gain compensator for imbalanced outputs of 
the two detectors.
The signals from high-speed photo detectors HPD1 and HPD2 were summed with a power combiner to produce DFI output 
signal, which was then monitored with a network analyzer (Anritsu MS4630B).
	
The DFI features of bi-directional MZIs were demonstrated with a transfer function measurement 
from noise/signal simulators to DFI output. 
A swept-sine noise signal was provided by the internal oscillator of the network analyzer and was applied to 
the EOM(s).
	\begin{figure}[ht]
	\begin{center}
	\includegraphics[width=9cm]{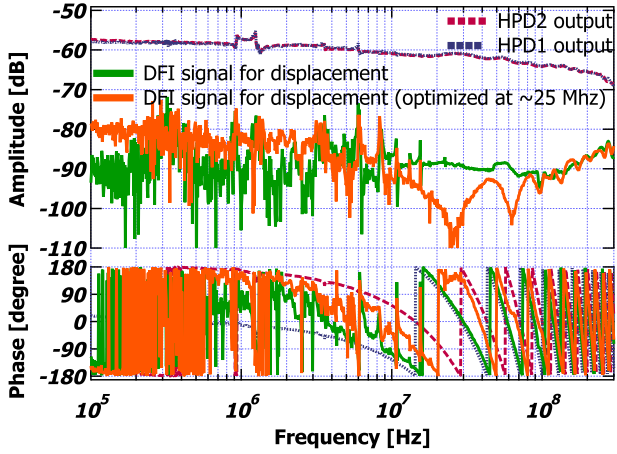}
	\caption{\label{fig3:disp}  (color online) 
	The response functions from displacement noise to output 
	signals. Dashed-red and dotted-blue curves are response to single detector HPD1 and HPD2, giving almost 
	identical response in amplitude while opposite in sign. Therefore summation of these two 
	signals gives cancellation of displacement noise.
	The suppression depends on a delicate tuning of the balance between two signals.
	Approximately 30\,dB of suppression can be seen in a wide frequency band from 10\,kHz to 
	300\,MHz in the green curve, meanwhile, the orange curve shows maximum suppression exceeding
	45\,dB at around 25\,MHz.}
	\end{center}
	\end{figure}	

The results for displacement noise suppression is shown in Fig.~\ref{fig3:disp}.
The transfer function of both detectors include the response functions of the simulating 
EOM, the photodetectors and optical and electronic phase delays from EOM through to the power
combiner. 
Both amplitude and phase for the two signals were tuned to match each other using optical 
attenuators and path lengths adjustments so that the displacement noise signal disappears in the DFI signal.
About 30\,dB of suppression of displacement noise was attained in a wide frequency 
region, while the maximum attained suppression was 45\,dB at a particular frequency band, 
achieved when we tuned for maximum suppression in that region.
Any imbalance between the two MZI signals determines the suppression ratio; it is believed that
the subtly different frequency response of our
two photodetectors, which cannot be compensated by simple optical attenuators or
path length tuning, was the limiting factor here.
	
	\begin{figure}[ht]
	\begin{center}
	\includegraphics[width=9cm]{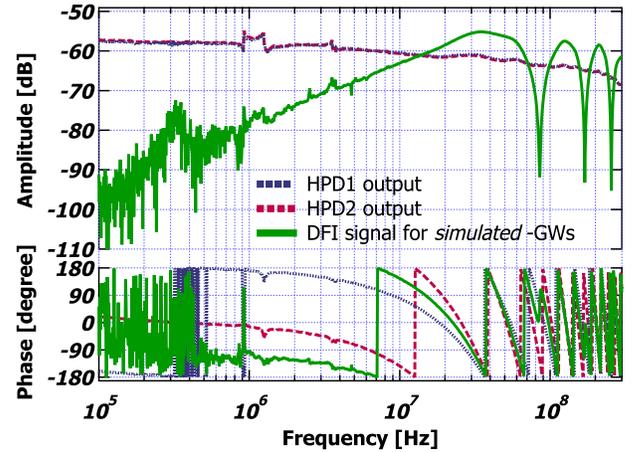}
	\caption{\label{fig4:gw} (color online) 
	The response functions from {\it simulated}-gw inputs to 
	output signals. 
	HPD1 and HPD2 response show almost the same response again in amplitude, whereas
	the phase delays in the higher frequency band are essentially different from 
	that of displacement noise, which is the very reason why the summed signal (DFI signal) 
	remains gw-sensitive as is shown in the solid-green curve.}
	\end{center}
	\end{figure}
\begin{figure}[h]
	\begin{center}
	\includegraphics[width=9cm]{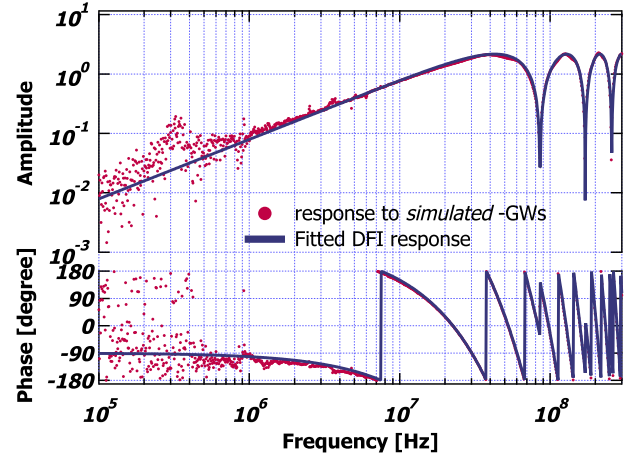}
	\caption{\label{fig5:smzi}  (color online) 
	The effects of the response functions of EOMs and HPDs  
	on the DFI signal (solid-green curve in Fig.~\ref{fig4:gw}) were compensated by using  
	the response function to single detector, shown in red dots. This corresponds to 
	an intrinsic DFI (bi-directional MZIs for this experiment) response function to a {\it simulated}
	gw signal. The solid curve is a fitted DFI response giving reasonable fitting parameters.}
	\end{center}
	\end{figure}

	
On the other hand, for the {\it simulated-}gw signal, the measured response function is 
shown in Fig.~\ref{fig4:gw}.
Single detector outputs show response function in amplitude similar to to that for displacement 
noise; however, phase responses are different which is what enables gw signal detection with DFI.
As the DFI signal still contains the response functions of the simulating EOM and photodetectors,
the measured transfer function was compensated with single detector data to remove them, 
then fitted with the theoretical function Eq.\eqref{sgw-res} as shown in Fig.~\ref{fig5:smzi}.
Fitted parameters, scale factor $\rm A=2.16$ (2 in theory) and $l_{1\alpha}-l_{1\beta}=3.50$ (3.55 in 
scale-measurement), agreed well with predicted values. 

\begin{acknowledgments}
We thank Albrecht R\"udiger and Koji Arai for helpful discussions and comments.
This research was partially supported by the Ministry of Education, Culture, Sports, Science and Technology under Grant-in-Aid 
for Scientific Research (B), 18340070, 2006. 
R.L.W. was supported by the U.S. National Science Foundation under Cooperative Agreement PHY-0107417.
Research of Y.C., A.P.\ and K.S.\ are supported by the Alexander von Humboldt Foundation's Sofja Kovalevskaja 
Programme (funded by the German Ministry of Education and Research). 
\end{acknowledgments}


\begin{thebibliography}{9}
\frenchspacing
\bibitem{TDI} M.~Tinto, F.B.~Estabrook and J.W.~Armstrong, Phys.\ Rev.\ D, {\bf 65}, 082003 (2002). 
\bibitem{KC1} S. Kawamura and Y. Chen, Phys. Rev. Lett. {\bf 93}, 211103 (2004). 
\bibitem{KC2} Y. Chen and S. Kawamura, Phys. Rev. Lett. {\bf 96}, 231102 (2006). 
\bibitem{KC3} Y. Chen {\it et al.},  Phys. Rev. Lett. {\bf 97}, 151103 (2006).
\end{thebibliography}

\end{document}